\documentclass[aps,pra,onecolumn,floatfix,showpacs,hyperref,nofootinbib,preprint]{revtex4}
\usepackage[dvips]{epsfig}
\usepackage{amsmath}
\usepackage{amssymb}
\usepackage{mathtools}
\usepackage{bbm}
\usepackage{hyperref}
\usepackage{xcolor}
\usepackage{color}
\usepackage{graphicx}
\usepackage{braket}
\usepackage{IEEEtrantools}
\usepackage{wasysym}

\DeclareFontFamily{OT1}{pzc}{}
\DeclareFontShape{OT1}{pzc}{m}{it}{<-> s * [1.10] pzcmi7t}{}
\DeclareMathAlphabet{\mathpzc}{OT1}{pzc}{m}{it}

\newcommand{\eq}[1]{\begin{equation}#1\end{equation}}
\newcommand{\seqa}[1]{\begin{subequations}\begin{IEEEeqnarray}{rCl}#1\end{IEEEeqnarray}\end{subequations}}
\newcommand{\eqa}[1]{\begin{IEEEeqnarray}{rCl}#1\end{IEEEeqnarray}}

\newcommand{\hsp}{\hspace{1pt}}

\newcommand{\re}[1]{\operatorname{Re}[#1]}
\newcommand{\im}[1]{\operatorname{Im}[#1]}

\graphicspath{ {./} {./Figures/} }

\begin{document}

\title{Dissipative Bose-Einstein condensation in contact with a thermal reservoir}

\author{S.\ Caspar}
\author{F.\ Hebenstreit}
\author{D.\ Mesterh\'{a}zy}
\author{U.-J.\ Wiese}
 \affiliation{Albert Einstein Center for Fundamental Physics, Institute for Theoretical Physics, University of Bern, 3012 Bern, Switzerland}

\begin{abstract}
 We investigate the real-time dynamics of open quantum spin-$1/2$ or hardcore boson systems on a spatial lattice, which are governed by a Markovian quantum master equation. 
 We derive general conditions under which the hierarchy of correlation functions closes such that their time evolution can be computed semi-analytically. 
 Expanding our previous work [\textit{Phys.\ Rev.\ A} {\bf 93} \textit{(2016) 021602}] we demonstrate the universality of a purely dissipative quantum Markov process that drives the system of spin-$1/2$ particles into a totally symmetric superposition state, corresponding to a Bose-Einstein condensate of hardcore bosons. 
 In particular, we show that the finite-size scaling behavior of the dissipative gap is independent of the chosen boundary conditions and the underlying lattice structure. 
 In addition, we consider the effect of a uniform magnetic field as well as a coupling to a thermal bath to investigate the susceptibility of the engineered dissipative process to unitary and nonunitary perturbations. 
 We establish the nonequilibrium steady-state phase diagram as a function of temperature and dissipative coupling strength. 
 For a small number of particles $N$, we identify a parameter region in which the engineered symmetrizing dissipative process performs robustly, while in the thermodynamic limit $N\rightarrow \infty$, the coupling to the thermal bath destroys any long-range order.
\end{abstract}



\maketitle

\section{Introduction}
\label{Sec:Introduction}

With the advent of ultracold gases and trapped ions as tunable quantum simulators, experiments are now in the position to investigate the real-time evolution of quantum many-body systems directly with engineered model Hamiltonians \cite{Jane:2003,Bloch:2012,Blatt:2012}.
Recent years have seen tremendous progress that promises new insights at the intersection of condensed matter physics, high energy physics, and beyond \cite{Buluta:2009,Georgescu:2014}.
While experiments have been very successful in probing the dynamics of quantum many-body systems, it is fair to say that to date our theoretical understanding remains incomplete. 
Thus, many important problems regarding nonequilibrium quantum dynamics as, e.g., the characterization of steady states or the calculation of transport properties far from equilibrium are still beyond our reach. 
There are several reasons for this disparity. While exact diagonalization provides us with rigorous results it is limited to systems consisting of a small number of particles. 
The density matrix renormalization group (DMRG) \cite{White:1992zz,Schollwock:2005zz} allows for the efficient simulation of real-time dynamics for a large variety of one-dimensional gapped quantum many-body systems \cite{Cazalilla:2002,Vidal:2004,Verstraete:2004,White:2004}. 
Nevertheless, its applicability remains largely limited to short times due to the growth of entanglement. 
Quantum Monte Carlo (QMC) methods are not limited by macroscopic quantum correlations, however, their application to nonequilibrium dynamics is hindered by a severe complex phase problem.

In light of these limitations, it is quite remarkable that a class of open quantum many-body models \cite{Breuer:2007} can be solved without approximations.
This is of particular interest since open quantum system with engineered couplings to an environment \cite{Poyatos:1996} have been proposed recently for the preparation of quantum states \cite{Verstraete:2009,Kastoryano:2011,Sweke:2013,DallaTorre:2013,Caballar:2014}, quantum simulation \cite{Sweke:2014,Sweke:2015,Zanardi:2016}, as well as quantum computing \cite{Verstraete:2009,Kliesch:2011,Pastawski:2011,Sinayskiy:2014}. 
Typically, the nonequilibrium steady state (NESS) of the engineered dissipative process, which defines a unique fixed point for the dynamics, is known by construction;\footnote{More complicated examples might be possible, e.g., where the late-time asymptotic behavior is characterized by a limit cycle.} examples include condensates of bosons or $\eta$ states of fermions \cite{Diehl:2008,Kraus:2008}, condensates of hardcore bosons or quantum spins \cite{Schindler:2013}, $d$-wave pairing states of fermions \cite{Diehl:2010,Yi:2012}, and various topologically ordered states \cite{Roncaglia:2010,Diehl:2011,Bardyn:2013tqa,Budich:2015}. 
So far, for most of these systems the real-time evolution leading to the final state is less well understood. 
It is known, however, that the dissipative contributions to the dynamics might in fact close the hierarchy of $n$-particle correlation functions such that a semi-analytic solution of interacting quantum systems becomes feasible. This observation has been exploited in a number of works \cite{Esposito:2005a,Esposito:2005b,Znidaric:2010,Eisler:2011,Horstmann:2013,Znidaric:2015,Huffman:2015} and general closure conditions for the hierarchy of correlation functions have been derived for bosonic and fermionic models \cite{Zunkovic:2014}.

In recent work \cite{Caspar:2016}, we investigated a protocol for dissipative generation of a Bose-Einstein condensate (BEC) \cite{Diehl:2008,Kraus:2008,Schindler:2013}. In particular, we considered a macroscopic system of $N$ spin-$1/2$ particles, where the time evolution of the reduced density matrix $\rho$ is governed by a Markovian quantum master equation \cite{Lindblad:1975,Gorini:1976,Breuer:2007}
\eq{\label{Eq:Lindblad}\frac{d}{dt} \rho = \boldsymbol{\mathcal{L}}\rho = -i [H,\rho] + \sum_{\lambda}\boldsymbol{\mathcal{L}}_{\lambda} \rho \ ,}
$\lambda = 1, \ldots , d_{N}^2 - 1$, where $d_N = 2^{-N}$ denotes the dimension of the $N$-particle Hilbert space.
The Lindbladian $\boldsymbol{\mathcal{L}} = -i[H,\,\cdot~] + \sum_{\lambda}\boldsymbol{\mathcal{L}}_{\lambda}$ is a linear map on the set of mixed states and the commutator part defines the system Hamiltonian $H$, which is Hermitian. 
The dissipative coupling to the environment is described by the non-commutator part $\sum_{\lambda}\boldsymbol{\mathcal{L}}_{\lambda}$ of $\boldsymbol{\mathcal{L}}$. 
Written in Lindblad form, the action of its constituents $\boldsymbol{\mathcal{L}}_{\lambda}$ on $\rho$ is expressed in terms of jump operators $L_{\lambda}$
\eq{\boldsymbol{\mathcal{L}}_{\lambda} \rho = \gamma_{\lambda} \left( L_{\lambda}\rho L_{\lambda}^\dagger-\frac{1}{2} \big\{L_{\lambda}^\dagger L_{\lambda},\rho\big\} \right) \ ,}
where the rate parameters $\gamma_{\lambda} \geq 0$ characterize the relative strength of the dissipative couplings. 
Eq.\ \eqref{Eq:Lindblad} represents an effective description of the system where the environment has been integrated out, yielding operators $\sqrt{\gamma_{\lambda}} L_{\lambda}$ that are expressed in terms of the local spin degrees of freedom $s_x^{\alpha}$, where $\alpha = 1,2,3$ denotes the spin index and $x$ labels distinct particles. 
Specifically, in Ref.\ \cite{Caspar:2016} only a single class of non-Hermitian jump operators was considered, which acts on pairs of particles, i.e., $L_{xy} = \tfrac{1}{2} (s_x^+ + s_y^+) (s_x^- - s_y^-)$, where $s_x^{\pm} = s_x^1 \pm i s_x^2$. In the absence of a Hamiltonian contribution ($H = 0$), we were able to show that this process yields a closed hierarchy of correlations, thereby allowing us to solve for the full real-time evolution of spin-spin correlation functions. Thus, the dynamics of dissipative Bose-Einstein condensation could be followed explicitly by studying the relaxation towards the final state $\rho_{\textrm{NESS}} = |\textrm{BEC}\rangle \langle\textrm{BEC}|$, satisfying $\boldsymbol{\mathcal{L}} \rho_{\textrm{NESS}} = 0$.

Here, we go beyond this first exploratory work and establish the generality of our approach, which allows us to investigate the susceptibility of the considered quantum Markov process to unitary and nonunitary perturbations. 
Specifically, we consider operators $L_{x}$ that act locally on particles at site $x$, as well as bilocal jump operators $L_{xy}$ that act on pairs of particles $(x,y)$, both in the presence and absence of a uniform magnetic field. 
While multilocal operators $L_{x_1 x_2 \cdots x_n}$, with $n > 2$, are conceivable and have been studied theoretically, e.g., in the context of steady states with nontrivial topology \cite{Weimer:2010,Bardyn:2012,Freeman:2014,Lang:2015}, local and bilocal operators appear to be sufficient to describe the phenomenology of engineered dissipative quantum spin-$1/2$ systems for BEC generation \cite{Barreiro:2011,Schindler:2013}.

The outline of this paper is as follows: 
In Sec.\ \ref{Sec:Closed hierarchies for correlation functions} we derive the closure conditions for hierarchies of correlation functions of quantum spin systems in the $s=1/2$ representation in the case of open Markovian dynamics governed by local and bilocal jump operators. 
In Sec.\ \ref{Sec:Dissipative cooling into a BEC} we discuss the dynamics of an engineered dissipative process that drives the system into a mixture of totally symmetric superposition states.
We describe the growth of long-range correlations in real time and consider the finite-size scaling of the dissipative gap. 
In Sec.\ \ref{Sec:Dissipative dynamics in the presence of competing unitary and nonunitary processes} we investigate the same dissipative process in the presence of thermal noise and a uniform magnetic field. 
Our results are summarized in the nonequilibrium steady-state phase diagram for the coupled model. 
We conclude with an outlook on applications and possible further developments of this work.

\section{Closed hierarchies for correlation functions}
\label{Sec:Closed hierarchies for correlation functions}

Throughout this work we consider quantum spin-$1/2$ systems on a regular $d$-dimensional lattice. 
The spin degrees of freedom $s_x^{\alpha} = \tfrac{1}{2} \sigma_x^{\alpha}$ are expressed in terms of Pauli matrices defined locally at site $x$, while the total spin operator is given by $S^\alpha = \sum_x s_x^\alpha$. 
The assumption of an underlying regular lattice structure is not essential, however, it allows us to exploit the symmetries of the lattice to simplify the problem and to solve for the dynamics of correlation functions of macroscopic quantum many-body systems.

In the following, we consider the dynamics of $n$-point correlation functions $\mathcal{O}_{z_1 z_2 \cdots z_n} = \operatorname{tr} \left\{ \rho \hsp O_{z_1 z_2 \cdots z_n } \right\}$, in particular, products of spin operators $O_{z_1 z_2 \cdots z_n } = \prod_{1\leq i \leq n} s_{z_i}$ (where spin indices $\alpha_i$ are omitted). 
Using Eq.\ \eqref{Eq:Lindblad} we derive the equation of motion
\eqa{\frac{d}{dt} \mathcal{O}_{z_1 z_2 \cdots z_n} &=& \operatorname{tr} \left\{ \rho \hsp \boldsymbol{\mathcal{L}}^{\boldsymbol{\ast}} O_{z_1 z_2 \cdots z_n} \right\} \nonumber\\
  &=& i \operatorname{tr} \left\{ \rho \left[H,O_{z_1 z_2 \cdots z_n }\right] \right\} + \frac{1}{2} \sum_{\lambda} \gamma_{\lambda} \operatorname{tr} \Big\{ \rho \Big( L_{\lambda}^\dagger\big[O_{z_1 z_2 \cdots z_n } ,L_{\lambda}\big]+\big[L_{\lambda}^\dagger, O_{z_1 z_2 \cdots z_n }\big]L_{\lambda} \Big) \Big\} \ , \nonumber\\[-3pt] && \label{Eq:AdjointLindblad}}
where we have introduced the dual (or adjoint) map $\boldsymbol{\mathcal{L}}^{\boldsymbol{\ast}} = i [H ,\,\cdot~] + \sum_{\lambda} \boldsymbol{\mathcal{L}}_{\lambda}^{\boldsymbol{\ast}}$ corresponding to the Lindbladian $\boldsymbol{\mathcal{L}}$. 
In general, Eq.\ \eqref{Eq:AdjointLindblad} does not close, i.e., the commutators on the right hand side typically induce operators of higher order. 
This leads to an infinite hierarchy of correlation functions which cannot be solved without truncating the coupled set of equations. Here, we seek conditions under which the opposite is true. 
That is, we derive conditions under which the hierarchy of correlation functions closes for a purely dissipative quantum Markov process ($H = 0$) defined in terms of local (Sec.\ \ref{SubSec:Local jump operators}) or bilocal jump operators (Sec.\ \ref{SubSec:Bilocal jump operators}). 
We discuss possible solutions to these conditions and return to the effect of unitary perturbations in Sec.\ \ref{SubSec:Hamiltonian contributions}.

\subsection{Local jump operators}
\label{SubSec:Local jump operators}

\begin{figure}[!t]
\includegraphics[width=0.8\textwidth]{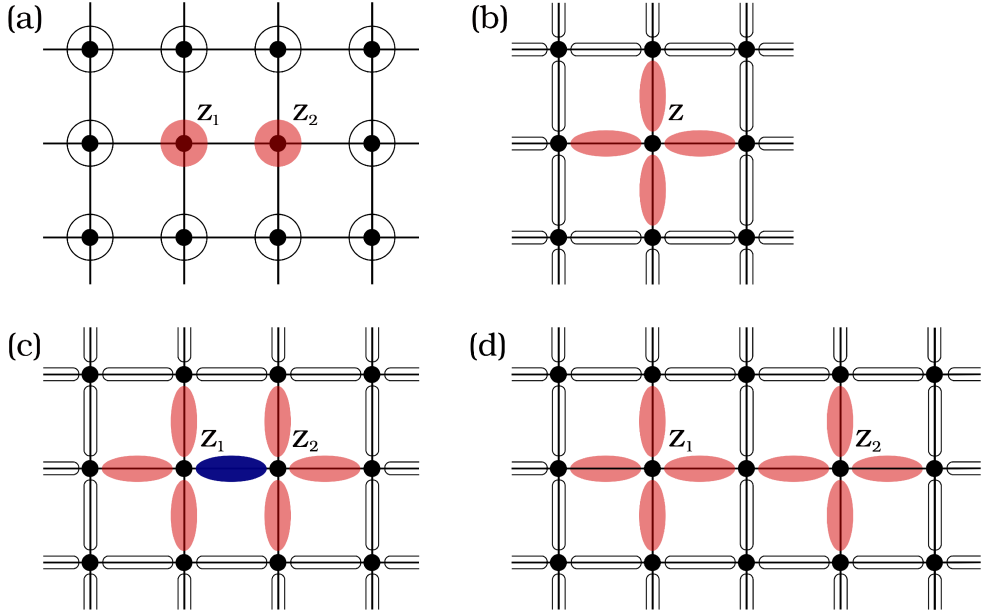}
\caption{[\textit{Color online}] Graphical illustration of terms that contribute to the purely dissipative ($H=0$) time evolution of correlation functions on a square lattice (with coordination number $n_c = 2 d$) for (a) local jump operators $L_x$  and  (b) -- (d) bilocal jump operators $L_{xy}$.
  (a) To determine the contributions to any $n$-point function $\mathcal{O}_{z_1 z_2 \cdots z_n}$, we sum over all insertions of $L_{z_i}$ ({\it red}) at lattice sites $z_1$, $z_2$, \ldots, $z_n$ and evaluate the trace [cf.\ Eq.\ \eqref{Eq:AdjointLindblad}]. This is illustrated above for the example of the two-point function $\mathcal{O}_{z_1 z_2}$ for which we show only the nonvanishing contributions.
  (b) To evaluate the time evolution of the local magnetization $\operatorname{tr} \{ \rho s_{z}\}$ we attach $n_c$ operators $L_{yz}$ ({\it red}) to the lattice site $z$ [cf.\ Eq.\ \eqref{Eq:EquationOfMotionLocalMagnetizationBilocalJumpOperator}].
  (c) For spin-spin correlation functions $\mathcal{O}_{z_1 z_2}$, where $z_1$ and $z_2$ are nearest-neighbor sites, we attach one bilocal jump operator that connects both sites ({\it blue}) while $n_c-1$ jump operators are assigned separately to $z_1$ and $z_2$ ({\it red}); see Eq.\ \eqref{Eq:AdjointLindbladBilocal}.
  (d) In the case of correlation functions $\mathcal{O}_{z_1 z_2}$ with nonadjacent sites $z_1$ and $z_2$, we attach $n_c$ bilocal jump operators to each site.}
\label{Fig:Lindblad}
\end{figure}

We start with the discussion of local operators $\sqrt{\gamma_x} L_x$ ($\gamma_x > 0$) to illustrate the derivation of closure conditions. 
Any local jump operator can be expressed in the form
\eq{\label{Eq:LocalJumpOperator}L_x = l^0 \mathbbm{1} + \sum_{\alpha} l^{\alpha} s_{x}^\alpha \ ,}
where the coefficients $l^0$ and $l^{\alpha}$, $\alpha = 1,2,3$, are defined up to an overall phase factor, which we may use to set $l^0\in\mathbbm{R}$ and $l^{\alpha}\in\mathbbm{C}$. 
Since $L_x$ acts locally and spin operators at different sites commute, we observe that the commutator terms in Eq.\ \eqref{Eq:AdjointLindblad} only contribute when $x = z_i$, $i = 1, \ldots, n$ (cf.\ Fig.\ \ref{Fig:Lindblad}a). 
Thus, the equation of motion for $n$-point correlation functions can be written as
\eq{\label{Eq:AdjointLindbladLocal}\frac{d}{dt} \mathcal{O}_{z_1 z_2 \cdots z_n} = \sum_{i=1}^{n} \operatorname{tr} \left\{ \rho \hsp\bigg(\prod_{j\neq i}^n s_{z_j} \bigg) \boldsymbol{\mathcal{L}}_{z_i}^{\boldsymbol{\ast}}s_{z_i} \right\} \ ,}
with the dual superoperator $\boldsymbol{\mathcal{L}}_x^{\boldsymbol{\ast}} = \gamma_x /2 \left(  L_x^{\dagger}\big[ ~\cdot\, ,L_x\big]+\big[L_x^\dagger, \,\cdot~\big]L_x \right)$, which maps the spin $s_z^{\alpha}$ at site $z = x$ onto the complete basis of the local spin-$1/2$ operator algebra, $\{ \mathbbm{1}, s^1_z, s^2_z, s^3_z \}$, and to zero otherwise ($z \neq x$).
We may express its action in the following form: 
\eq{\label{Eq:AdjointLindbladLocalParametrization}\frac{1}{\gamma_x} \boldsymbol{\mathcal{L}}_x^{\boldsymbol{\ast}} \hsp s_z^{\alpha} = \Big( m^{\alpha} \mathbbm{1} + \sum_{\beta} m^{\alpha\beta} s_z^{\beta} \Big) \delta_{xz} \ ,}
where the additional factor $1/\gamma_x$ is introduced to render the right hand side dimensionless. 
The coefficients $m^{\alpha}$ and $m^{\alpha\beta}$ are obtained by using the parametrization of the local jump operator \mbox{[cf.\ Eq.\ \eqref{Eq:LocalJumpOperator}]} to evaluate the action of $\boldsymbol{\mathcal{L}}_x^{\boldsymbol{\ast}}$; we find
\seqa{m^\alpha &=& \frac{1}{4} \sum_{\beta,\gamma} \epsilon^{\alpha \beta \gamma} \im{\bar{l}^{\beta} l^{\gamma}} \ ,\\
     m^{\alpha\beta} &=& \frac{1}{2} \,\bigg( \!\re{\bar{l}^{\alpha} l^{\beta}} - \delta^{\alpha\beta} \sum_{\gamma} |l^{\gamma}|^2 \bigg) + \hsp l^0 \sum_{\gamma}\epsilon^{\alpha \beta \gamma}\im{l^{\gamma}} \ ,}
and the bar, e.g., $\bar{l}^{\alpha}$, denotes complex conjugation.
We observe that the equation of motion closes, i.e., the equations of motion for $n$-point correlation functions do not couple to higher order correlations, independent of the particular form of the local jump operators $L_x$. This does not hold true for generic jump operators that act on multiple sites, as we illustrate in the next section.

\subsection{Bilocal jump operators}
\label{SubSec:Bilocal jump operators}

A large number of suggested protocols for dissipative state generation rely on Lindblad dynamics induced by jump operators that act on pairs of particles (see, e.g.\ Ref.\ \cite{Bardyn:2013tqa}). It is therefore interesting to ask under what conditions the real-time dynamics can be solved either exactly or semi-analytically. Here, we restrict ourselves to bilocal operators $\sqrt{\gamma_{xy}}L_{xy}$ ($\gamma_{xy} > 0$) that act isotropically on adjacent lattice sites $x$ and $y$ (which we denote by $\langle x,y\rangle$). Although the assumptions of isotropy and nearest-neighbor couplings are both reasonable, they are not necessary for the following derivations. They serve merely to make the following arguments more transparent and to simplify the discussion.

For isotropic systems, any jump operator $L_{xy}$ is either symmetric ($L_{xy} = L_{yx}$) or antisymmetric ($L_{xy} = -L_{yx}$) under the interchange of $x$ and $y$. By making use of this fact, we may reduce the number of coefficients that parametrize any bilocal operator $L_{xy}$ and thereby provide explicit closure conditions for such operators.

\subsubsection{Symmetric jump operators}
\label{SubSubSec:Symmetric jump operators}

We employ the following parametrization for symmetric operators
\eq{\label{Eq:BilocalJumpSymmetric}L_{xy} = l^0\mathbbm{1} + i \sum_{\alpha} l^\alpha ( s_x^{\alpha} + s_y^{\alpha}) + \sum_{\alpha, \beta} l^{(\alpha\beta)} s_x^\alpha s_y^\beta \ ,}
where we have defined the symmetrization of indices by $l^{(\alpha\beta)} = \frac{1}{2} (l^{\alpha\beta} + l^{\beta\alpha})$ and the coefficients can be chosen as $l^0 \in \mathbb{R}$ and $l^\alpha$, $l^{\alpha\beta}\in \mathbb{C}$. The factor of $i$ is introduced for later convenience. Specific examples for symmetric jump operators, whose purely dissipative dynamics has been investigated recently via a novel Monte Carlo method \cite{Banerjee:2014,Hebenstreit:2015,Banerjee:2015}, are the singlet projection operator
\eqa{P_{xy}^{\,\textrm{s}} = \frac{1}{4}\mathbbm{1}-\sum_{\alpha}s_{x}^{\alpha}s_{y}^{\alpha} \ ,}
and triplet projection operator
\eq{P_{xy}^{\,\textrm{tr}} = \frac{3}{4}\mathbbm{1}+\sum_{\alpha}s_{x}^{\alpha}s_{y}^{\alpha} \ ,}
respectively.

\subsubsection{Antisymmetric jump operators}
\label{SubSubSec:Antisymmetric jump operators}

In the case of antisymmetric operators, we may choose the following parametrization
\eq{\label{Eq:BilocalJumpAntisymmetric}L_{xy} = i \sum_{\alpha} l^{\alpha} ( s_{x}^{\alpha} - s_y^{\alpha}) + \sum_{\alpha, \beta} l^{[\alpha\beta]} s_{x}^\alpha s_{y}^\beta \ ,}
where $l^{[\alpha\beta]} = \frac{1}{2} (l^{\alpha\beta} - l^{\beta\alpha})$ and the parameters $l^{\alpha}$ and $l^{\alpha\beta}$ are complex valued in general but due to the phase ambiguity, we may choose any one of these parameters to be real. The non-Hermitian operator
\eq{\label{Eq:NonHermitianJumpOperator}Q_{xy} = \frac{1}{2} \left( s_x^{+} + s_y^{+} \right) \left( s_x^{-} - s_y^{-} \right) \ ,}
which provides a mechanism for dissipative cooling into a BEC \cite{Caspar:2016,Diehl:2008,Kraus:2008,Schindler:2013} falls into this class of operators.

\subsubsection{Closure conditions}
\label{SubSubSec:Closure conditions}

Before we go on to discuss the closure conditions for $n$-point spin correlation functions, we derive the equation of motion for the local magnetization $\mathcal{O}_z = \operatorname{tr}\{ \rho \hsp s_z\}$ and investigate under which conditions it decouples from higher order correlation functions.

Evaluating Eq.\,\eqref{Eq:AdjointLindblad} for $n = 1$ we find that the commutator terms on the right hand side contribute only when jump operators $L_{yz}$ are attached to the site $z$ (cf.\ Fig.\ \ref{Fig:Lindblad}b). In fact, these contributions have to be summed over all nearest-neighbor pairs $\langle y, z\rangle$, so that 
\eq{\label{Eq:EquationOfMotionLocalMagnetizationBilocalJumpOperator}\frac{d}{dt} \mathcal{O}_z = \sum_{y\hsp |\langle y,z\rangle} \operatorname{tr} \left\{\rho \hsp\boldsymbol{\mathcal{L}}_{yz}^{\boldsymbol{\ast}} \hsp s_z \right\} \ .}
The dual map $\boldsymbol{\mathcal{L}}_{xy}^{\boldsymbol{\ast}}$ is defined in terms of bilocal operators, i.e., $\boldsymbol{\mathcal{L}}_{xy}^{\boldsymbol{\ast}} = \gamma_{xy} /2 \left(  L_{xy}^{\dagger}\big[ ~\cdot\, ,L_{xy}\big]+\big[L_{xy}^\dagger, \,\cdot~\big]L_{xy} \right)$ and its action on single spin operators can be expressed in the form 
\eq{\label{Eq:BilocalJumpCommutator}\frac{1}{\gamma_{xy}} \boldsymbol{\mathcal{L}}_{xy}^{\boldsymbol{\ast}} \hsp s_z^{\alpha} = \bigg( m^{\alpha}\mathbbm{1} + \sum_{\beta} \left[ m_x^{\alpha\beta} s_x^{\beta} + m_y^{\alpha\beta} s_y^{\beta} \right] + \sum_{\beta,\gamma} m^{\alpha\beta\gamma} s_x^{\beta} s_y^{\gamma} \bigg) \left( \delta_{xz} + \delta_{yz} \right) \ .}
Note that $\boldsymbol{\mathcal{L}}_{xy}^{\boldsymbol{\ast}}$ is invariant under the interchange of sites $x$ and $y$.

Clearly, the equation of motion for the one-point function closes only when the contributions $\sim s_x s_y$ in Eq.\ \eqref{Eq:BilocalJumpCommutator} vanish. 
Hence, to derive the closure conditions for bilocal operators $L_{xy}$, we require that $m^{\alpha\beta\gamma} = 0$, which yields a set of $27$ equations. 
Due to the distinct parametrization of the symmetric and antisymmetric jump operators, we discuss both cases separately:

\begin{itemize}

\item[1)\,] We obtain the following set of closure conditions for symmetric jump operators
  [cf.\ Eq.\ \eqref{Eq:BilocalJumpSymmetric}]
  \seqa{\label{Eq:BilocalClosureSymmetricBegin}
    \im{\bar{l}^{\alpha}a^{\alpha}} &=& -\frac{1}{2}\left(\im{\bar{l}^{\alpha}a^{\beta}}+\im{\bar{l}^{\alpha}a^{\gamma}}+\im{\bar{l}^{\beta}b^{\gamma}}+\im{\bar{l}^{\gamma}b^{\beta}}\right) \ , \quad \alpha\neq \beta\neq \gamma \ , \qquad \\
    \im{\bar{l}^{\alpha}l^{\beta}} &=& \frac{1}{4}\sum_{\gamma}\epsilon^{\alpha\beta\gamma}\left(\im{\bar{l}^{\gamma}a^{\alpha}}+\im{\bar{l}^{\gamma}a^{\beta}}-\im{\bar{l}^{\alpha}b^{\beta}}-\im{\bar{l}^{\beta}b^{\alpha}}\right) \ ,\\
    \im{\bar{a}^{\alpha}a^{\beta}} &=& 2\sum_{\gamma}\epsilon^{\alpha\beta\gamma}\left(\im{\bar{l}^{\alpha}b^{\alpha}}+\im{\bar{l}^{\beta}b^{\beta}}\right) \ ,\\
    \im{\bar{a}^{\alpha}b^{\beta}} &=& -2\sum_{\gamma}\epsilon^{\alpha\beta\gamma}\im{\bar{l}^{\alpha}b^{\gamma}}-\delta^{\alpha\beta}\sum_{\gamma,\delta}\epsilon^{\alpha\gamma\delta}\left(\im{\bar{l}^{\alpha}a^{\gamma}}+\im{\bar{l}^{\gamma}b^{\delta}}\right) \ ,\\
    \im{\bar{b}^{\alpha}b^{\beta}} &=& \sum_{\gamma}\epsilon^{\alpha\beta\gamma}\left(\im{\bar{l}^{\gamma}a^{\alpha}}+\im{\bar{l}^{\gamma}a^{\beta}}+\im{\bar{l}^{\alpha}b^{\beta}}+\im{\bar{l}^{\beta}b^{\alpha}}\right) \ ,\\
    l^0\im{a^{\alpha}} &=& -\frac{1}{2}\sum_{\beta,\gamma}\epsilon^{\alpha\beta\gamma}\im{\bar{l}^{\beta}b^{\beta}} \ ,\\
    l^0\im{b^{\alpha}} &=& -\frac{1}{4}\sum_{\beta,\gamma}\epsilon^{\alpha\beta\gamma}\left(\im{\bar{l}^{\alpha}a^{\beta}}-\im{\bar{l}^{\beta}b^{\gamma}}\right) \label{Eq:BilocalClosureSymmetricEnd} \ ,}
where we have introduced the coefficients $a^{\alpha} = l^{\alpha\alpha}$, with $\alpha = 1,2,3$, as well as $b^1 = l^{(23)}$, $b^2 = l^{(31)}$, and $b^3 = l^{(12)}$. From these relations we see that a $10$-parameter family of symmetric jump operators can be found (with $l^0$, $l^{\alpha}$, $a^{\alpha}$, $b^{\alpha}\in\mathbbm{R}$), for which the above conditions are fulfilled and the local magnetization decouples from the higher order $n$-point correlation functions ($n \geq 2$). Both the singlet and triplet projection operators $P^{\,\textrm{s}}_{xy}$ and $P^{\,\textrm{tr}}_{xy}$ belong to this class of operators.

\item[2)\,] For the antisymmetric jump operators
  [cf.\ Eq.\ \eqref{Eq:BilocalJumpAntisymmetric}] we obtain
\eq{\label{Eq:BilocalClosureAntisymmetric}
  \im{\bar{l}^{\alpha} k^\beta} = \im{\bar{l}^{\alpha} l^{\beta}} = \im{\bar{k}^\alpha k^\beta} = 0 \ , }
where $k^1 = l^{[23]}$, $k^2 = l^{[31]}$, and $k^3 = l^{[12]}$.
We find a 6-parameter family of antisymmetric operators $L_{xy}$ (with $l^{\alpha}$, $k^{\alpha} \in \mathbb{R}$) for which the equation of motion of the one-point function closes.
Clearly, the non-Hermitian operator $Q_{xy}$ belongs to this class.

\end{itemize}

It remains to be shown that Eqs.\ \eqref{Eq:BilocalClosureSymmetricBegin} -- \eqref{Eq:BilocalClosureSymmetricEnd} and Eq.\ \eqref{Eq:BilocalClosureAntisymmetric} are sufficient to close the hierarchy for arbitrary $n$-point correlation functions. 
In the following we demonstrate this explicitly for two-point functions, but similar arguments also apply to correlation functions of higher order. 
The equation of motion for $\mathcal{O}_{z_1z_2} = \operatorname{tr}\{\rho \hsp s_{z_1}s_{z_2}\}$ is given by
\eq{\label{Eq:AdjointLindbladBilocal}\frac{d}{dt} \mathcal{O}_{z_1z_2} = \sum_{i=1}^{2} \sum_{\substack{y \hsp|\langle y,z_i\rangle\\ \, y\neq z_{j\neq i}}} \operatorname{tr}\left\{ \rho \hsp \boldsymbol{\mathcal{L}}_{yz_i}^{\boldsymbol{\ast}} (s_{z_1} s_{z_2}) \right\} + \delta_{\langle z_1,z_2\rangle} \operatorname{tr} \left\{ \rho \hsp \boldsymbol{\mathcal{L}}_{z_1z_2}^{\boldsymbol{\ast}} (s_{z_1} s_{z_2}) \right\} \ ,}
where $\delta_{\langle z_1,z_2\rangle}$ is equal to one only if $z_1$ and $z_2$ are nearest-neighbor sites. 
Note that the contributions on the right hand side can be simplified by using, e.g.,
\eq{\label{Eq:TrilocalJumpCommutator}\boldsymbol{\mathcal{L}}_{yz_1}^{\boldsymbol{\ast}} (s_{z_1}^{\alpha_1} s_{z_2}^{\alpha_2}) = (1 - \delta_{yz_2}) \hsp s_{z_2}^{\alpha_2} \boldsymbol{\mathcal{L}}_{y z_1}^{\boldsymbol{\ast}} s_{z_1}^{\alpha_1} + \delta_{yz_2} \boldsymbol{\mathcal{L}}_{z_1z_2}^{\boldsymbol{\ast}} (s_{z_1}^{\alpha_1} s_{z_2}^{\alpha_2}) \ .}
Imposing the closure conditions for the local magnetization, i.e., Eqs.\ \eqref{Eq:BilocalClosureSymmetricBegin} -- \eqref{Eq:BilocalClosureSymmetricEnd} for symmetric and Eq.\ \eqref{Eq:BilocalClosureAntisymmetric} for antisymmetric jump operators, respectively, we observe that the right hand side of Eq.\ \eqref{Eq:TrilocalJumpCommutator} can be expressed in terms of either single spin or pairs of spin operators.
As a consequence, the equation of motion for the two-point function $\mathcal{O}_{z_1z_2}$ [cf.\ Eq.\ \eqref{Eq:AdjointLindbladBilocal}] decouples from higher order correlation functions. We have therefore shown that the closure conditions for one-point functions are indeed sufficient to close the equation of motion for two-point correlation functions. These arguments can be trivially generalized to arbitrary $n$-point correlation functions.

\subsection{Hamiltonian contributions}
\label{SubSec:Hamiltonian contributions}

We briefly comment on the situation when $H \neq 0$, inquiring under what conditions the hierarchy for $n$-point correlation functions closes. 
To this end, we assume a general Hamiltonian that takes the following form
\eq{H = \sum_{x}\sum_{\alpha}h^{\alpha}s_{x}^{\alpha} +\sum_{\langle x,y\rangle}\sum_{\alpha,\beta}J^{\alpha\beta}s_{x}^{\alpha}s_{y}^{\beta}\ ,}
with real-valued coefficients $h^\alpha$, $J^{\alpha\beta} = J^{\beta\alpha}$.
We evaluate the first term on the right hand side in Eq.\ \eqref{Eq:AdjointLindblad}, which yields
\eq{i[H,s_{x}^{\alpha}] = \sum_{\beta,\gamma} \epsilon^{\alpha\beta\gamma} h^{\beta} s_{x}^{\gamma}+\sum_{y\hsp |\langle y,x\rangle}\sum_{\beta,\gamma,\delta}\epsilon^{\alpha\beta\delta} J^{\beta\gamma} s_{y}^\gamma s_{x}^\delta \ .}
Calculating the ensemble average, we see that the second term on the right hand side will induce a dependence on higher order correlation functions. 
Thus, to close the hierarchy we require that $J^{\alpha\beta} = 0$. Accordingly, we will only allow for local Hamiltonian contributions in the following.

\section{Dissipative cooling into a BEC}
\label{Sec:Dissipative cooling into a BEC}

In previous work \cite{Caspar:2016}, we studied the purely dissipative dynamics of a many-body quantum spin system on a hypercubic $d$-dimensional lattice with periodic boundary conditions. The considered dissipative process is distinguished by non-Hermitian jump operators that lead to the growth of correlations and the generation of macroscopic order in the final state. We discussed the dependence of the dissipative gap on the system size and found a novel nontrivial scaling behavior as a function of the system size $N=L^d$, where $L$ denotes the linear extent.
In this section, we briefly summarize our main results and comment on the universality of this process with respect to the chosen lattice structure and boundary conditions of the problem.
In Sec.\ \ref{Sec:Dissipative dynamics in the presence of competing unitary and nonunitary processes}, we augment this investigation by including the effect of thermal noise and studying the stability of the dissipative process.

\subsection{Non-Hermitian jump operator and nonequilibrium steady state}
\label{SubSec:Non-Hermitian jump operator and nonequilibrium steady state}

We consider a quantum Markov system of spin-$1/2$ particles that is driven uniformly on the lattice by the application of non-Hermitian operators $\sqrt{\gamma} Q_{xy}$, where
\eq{Q_{xy} = \frac{1}{2}(s_{x}^{+}+s_{y}^{+})(s_x^{-}-s_{y}^{-}) = \frac{1}{2}(s_{x}^{3}-s_{y}^{3})+i(s_{x}^{1}s_{y}^{2}-s_{x}^{2}s_{y}^{1}) \ ,}
which acts on adjacent lattice sites $\langle x,y\rangle$. The associated dissipative coupling, denoted by $\gamma$, is independent of the particular pair of particles $\langle x,y\rangle$. The operator $Q_{xy}$ maps any two-particle spin-singlet state to the spin triplet, while conserving the total spin projection $S^3 = \sum_x s_x^3$ along the quantization axis, and annihilates the spin triplet state, i.e., $Q_{xy}^2 = 0$.

The final state $\rho_{\textrm{NESS}}$ of the time evolution is determined by the fixed point of the dynamic map
\eq{\boldsymbol{\mathcal{L}} \rho_{\textrm{NESS}} = 0 \ .}
This state is unique for a given initial state and corresponds to an ensemble of totally symmetric superposition states \cite{Diehl:2008}. By virtue of the quantum spin-$1/2$ to hardcore boson mapping \cite{Matsubara:1956}, this dissipative process can also be seen as a symmetric delocalization of hardcore bosons over adjacent sites, with a BEC of hardcore bosons as the resulting final state.

\subsection{Evolution equations for correlation functions}
\label{SubSec:Evolution equations for correlation functions}

In Sec.\ \ref{SubSec:Bilocal jump operators} we have seen that the hierarchy of correlation functions closes for the non-Hermitian jump operator $Q_{xy}$.
Here, we derive the explicit form for the equations of motion of one- and two-point functions when the dynamics is governed only by the dissipative process $\sqrt{\gamma} Q_{xy}$ on nearest neighbor sites, neglecting the effect of Hamiltonian contributions.

\subsubsection{Local magnetization}
\label{SubSubSec:Local magnetization}

We begin with the time evolution equation for the local magnetization ${\scriptstyle\mathcal{S}}_x^{\alpha} = \operatorname{tr}\left\{ \rho \hsp s_x^{\alpha}\right\}$. To derive the equation of motion, we use the relation
\eq{Q_{xy}^{\dagger} [s_x^{\alpha},  Q_{xy}] + [Q_{xy}^{\dagger}, s_x^{\alpha}] Q_{xy} = \frac{1}{2}\big(s_{y}^{\alpha}-s_{x}^{\alpha}\big)\ ,}
and sum over all nearest-neighbor pairs $\langle x,y\rangle$, keeping $x$ fixed. Calculating the trace over the density matrix $\rho$ (using Eq.\ \eqref{Eq:AdjointLindblad}) we obtain the diffusion equation
\eq{\frac{d}{d \tau}{\scriptstyle\mathcal{S}}_{x}^{\alpha} = \frac{1}{4}\Delta_{x}{\scriptstyle\mathcal{S}}_{x}^{\alpha} \ .}
The dimensionless time variable $\tau = \gamma\hsp t$ is introduced for later convenience, as well as the discretized Laplacian $\Delta_x$, i.e., $\Delta_x \delta_{xy} = \sum_{\mu =1}^{d} \left(\delta_{(x+\hat{\mu})y} - 2\delta_{xy} + \delta_{(x-\hat{\mu})y}\right)$, where the lattice constant is set to one, and $\hat{\mu}$ denotes the unit vector in the $\mu$ direction on the regular spatial lattice.

\subsubsection{Spin-spin correlation functions}
\label{SubSubSec:Spin-spin correlation functions}

To follow the growth of long-range correlations we consider the ensemble average of bilocal spin operators $s_x^{\alpha}s_y^{\beta}$. In particular, we consider $C_{xy} = s_{x}^{+}s_{y}^{-}+s_{x}^{-}s_{y}^{+}$ and $D_{xy} = s_{x}^{3}s_{y}^{3}$, for which we define the expectation values $\mathcal{C}_{xy} = \operatorname{tr} \left\{ \rho\hsp C_{xy} \right\}$ and $\mathcal{D}_{xy} = \operatorname{tr} \left\{ \rho\hsp D_{xy} \right\}$. Note that similar to the local magnetization, we will use calligraphic fonts to denote ensemble-averaged quantities in the following. The adjoint map $\boldsymbol{\mathcal{L}}_{xz}^{\boldsymbol{\ast}}$ corresponding to the non-Hermitian jump operator $Q_{xz}$ maps the operators $C_{xy}$ and $D_{xy}$ ($x\neq y$) to
\seqa{Q_{xz}^{\dagger} [C_{xy},  Q_{xz}] + [Q_{xz}^{\dagger}, C_{xy}] Q_{xz} &=& \frac{1}{2} (1 - \delta_{yz})(C_{yz}-C_{xy}) + \delta_{yz} (C_{xx} - 2 C_{xy} - 4 D_{xy}) \ , \qquad \\
 Q_{xz}^{\dagger} [D_{xy},  Q_{xz}] + [Q_{xz}^{\dagger}, D_{xy}] Q_{xz} &=& \frac{1}{2} (1 - \delta_{yz}) (D_{yz}-D_{xy}) \ ,}
while the diagonal elements $C_{xx} = 4D_{xx} = \mathbbm{1}$ are mapped to zero.
The time evolution equation for the expectation values $\mathcal{C}_{xy}$ and $\mathcal{D}_{xy}$ are obtained by summing over all nearest-neighbor pairs of $x$ and $y$, and averaging over $\rho$ [cf.\ Eq.\ \eqref{Eq:AdjointLindbladBilocal}]. For $x\neq y$, we obtain
\seqa{\label{Eq:EvolutionEquation2PtCorrelationsDissipativeBegin}
 \frac{d}{d\tau}\mathcal{C}_{xy} &=& \frac{1}{4}\left(\Delta_x+\Delta_y\right)\mathcal{C}_{xy}-\frac{1}{2}\delta_{\langle x,y\rangle}\left(\mathcal{C}_{xy}+4\mathcal{D}_{xy}\right) \ ,\\
 \label{Eq:EvolutionEquation2PtCorrelationsDissipativeEnd}
 \frac{d}{d\tau}\mathcal{D}_{xy} &=& \frac{1}{4}\left(\Delta_x+\Delta_y\right)\mathcal{D}_{xy}+\frac{1}{2}\delta_{\langle x,y\rangle}\left(\mathcal{D}_{xy}-\mathcal{D}_{xx}\right) \ ,}
while the diagonal terms $\mathcal{C}_{xx} = 4\mathcal{D}_{xx} = 1$ are constant in time.
Given initial data for the two-point correlation functions, the above first-order system of differential equations
\eq{\label{Eq:EvolutionEquation2PtCorrelationsDissipative}
 \frac{d}{d\tau}\begin{pmatrix} \mathcal{C}_{xy} \\ \mathcal{D}_{xy} \end{pmatrix} = \mathcal{M} \begin{pmatrix} \mathcal{C}_{xy}\\ \mathcal{D}_{xy} \end{pmatrix} \ , }
can be solved explicitly via matrix diagonalization, where $\mathcal{M}$ denotes the linear differential operator that is determined by Eq.\ \eqref{Eq:EvolutionEquation2PtCorrelationsDissipativeBegin} and \eqref{Eq:EvolutionEquation2PtCorrelationsDissipativeEnd}.
Its solutions can be expressed in terms of a linear combination of exponential functions, whose characteristic decay rates are determined by the nonvanishing eigenvalues of $\mathcal{M}$ (see Sec.\ \ref{SubSec:Dynamics of purely dissipative cooling}).

We exploit spatial translation invariance in the following to explicitly solve the linear system of equations \eqref{Eq:EvolutionEquation2PtCorrelationsDissipative}. Accordingly, we may characterize the time evolution of spin-spin correlation functions in momentum space, e.g., 
\eq{\mathcal{C}_p=\frac{1}{N^2}\sum_{x,y}{e^{i \sum_{\mu} p_{\mu}(x-y)_{\mu}}\, \mathcal{C}_{xy} } \ , }
with $p_{\mu} = 2\pi n_{\mu}/L$ and $n_{\mu}\in\{0,\ldots,L-1\}$. The zero-momentum component $\mathcal{C}_{p=0}$ corresponds to the condensate fraction, which indicates the buildup of long-range correlations in the system.

\subsection{Dynamics of purely dissipative cooling}
\label{SubSec:Dynamics of purely dissipative cooling}

\begin{figure}[!t]
\includegraphics[width=\textwidth]{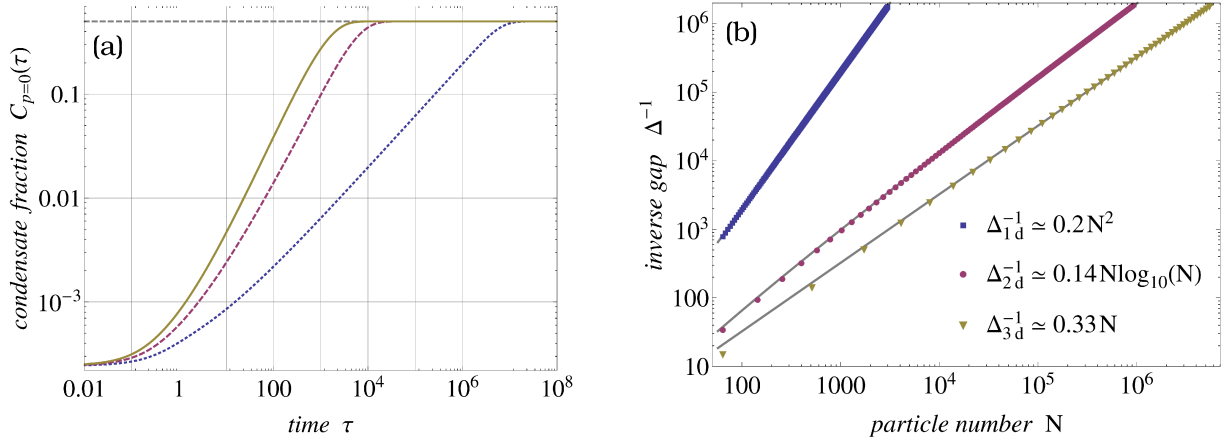}
\caption{[\textit{Color online}]
 (a) Time evolution of the condensate fraction $\mathcal{C}_{p=0}(\tau)$ for a fixed system size $N = 4096$ on a $d = 1$ ({\it dotted}), $d = 2$ square ({\it dashed}), and $d = 3$ primitive cubic ({\it solid}) lattice. The dashed horizontal line denotes the exact asymptotic value $\mathcal{C}_{p=0}^{\infty} = ( N + 1 ) / (2N)$.
 (b) Inverse dissipative gap $\Delta^{-1}$ evaluated on a regular lattice with periodic boundary conditions in $d = 1$ ({\it squares}), $d = 2$ ({\it dots}), and $d = 3$ ({\it triangles}) dimensions as a function of the system size $N$, cf. also Fig.\ 2 in \cite{Caspar:2016}.}
\label{Fig:DissipativeGap}
\end{figure}

As an example, we illustrate the time evolution starting from an incoherent thermal ensemble at infinite temperature
\eq{\rho(0) = d_N^{-1} \hsp \mathbbm{1} \ .}
For this initial state, the spin-spin correlation functions are easily evaluated, i.e.,
\eq{\mathcal{C}_{xy}(0) = 4\mathcal{D}_{xy}(0) = \delta_{xy} \ ,}
indicating the absence of long-range correlations in the system. 
Also the final state can be found explicitly,
\eq{\label{Eq:InfiniteTemperatureFinalEnsemble}\lim_{\tau\to\infty}\rho(\tau) = d_{N}^{-1}\sum_{n = 0}^N {N \choose n} |D(N,n) \rangle \langle D(N,n) | \ ,}
which is expressed in terms of totally symmetric Dicke states, $|D(N,n) \rangle = |N/2, -N/2 + n\rangle$, which are simultaneous eigenstates of $\sum_{\alpha} \big(S^{\alpha}\big)^2$ and $S^3$, respectively.
Using Eq.\ \eqref{Eq:InfiniteTemperatureFinalEnsemble} we determine the asymptotic values for the correlation functions
\seqa{\label{Eq:AsymptoticCorrelation}\mathcal{C}_{xy}^{\infty} &=& \lim_{\tau\to\infty}\mathcal{C}_{xy}(\tau) = \frac{1}{2}(1+\delta_{xy}) \ ,\\
  \mathcal{D}_{xy}^{\infty} &=& \lim_{\tau\to\infty}\mathcal{D}_{xy}(\tau) = \frac{1}{4} \delta_{xy} \ .}

To solve for the full real-time evolution we diagonalize Eq.\ \eqref{Eq:EvolutionEquation2PtCorrelationsDissipative} numerically. 
Our results are shown in Fig.\ \ref{Fig:DissipativeGap}a, where we display the time evolution of the condensate fraction $\mathcal{C}_{p=0}(\tau)$. 
The timescale on which the asymptotic regime is reached, i.e., $\mathcal{C}_{p = 0} > \mathcal{C}_{p = 0}^{\infty} e^{-\epsilon}$, \mbox{$0 < \epsilon \ll 1$}, depends strongly on dimension. 
We may understand this behavior by inspecting the spectrum of the linear operator $\mathcal{M}$ [see Eq.\ \eqref{Eq:EvolutionEquation2PtCorrelationsDissipative}]. 
The corresponding eigenvalues $\lambda \in \operatorname{Spec} \mathcal{M}$ take only nonpositive values and the mode corresponding to the eigenvalue
\eq{\Delta = -\max_{\lambda} \operatorname{Re}{ \lambda } > 0 \ ,}
which is denoted as dissipative gap in the following, dominates the asymptotic approach towards the final steady state.
Interestingly, we find by numerically diagonalizing the full Lindbladian $\boldsymbol{\mathcal{L}}$ on small system sizes that the value of the smallest negative eigenvalue of the linear operator $\mathcal{M}$ is identical to the dissipative gap of the Lindbladian $\boldsymbol{\mathcal{L}}$.
It seems that the spin-spin correlation functions capture the slowest mode of the full Lindbladian $\boldsymbol{\mathcal{L}}$ for the system under consideration.

For the purely dissipative process governed by the jump operators $Q_{xy}$, the coupling $\gamma$ can be scaled out, so that the total number of particles $N = L^d$ provides the only scale in the problem. Thus, the dissipative gap is a function of the system size and we may inquire about its asymptotic scaling properties, i.e., $\Delta\sim L^{-z}$, for sufficiently large $L$.
We remark that the finite-size scaling of $\Delta$ has been studied in detail for various one-dimensional bosonic and fermionic systems \cite{Esposito:2005a,Esposito:2005b,Prosen:2008,Znidaric:2011,Eisler:2011,Cai:2013,Horstmann:2013,Zunkovic:2014,Znidaric:2015,Crawford:2015}. However, here we observe that the considered purely dissipative quantum spin system exhibits a highly nontrivial scaling behavior that depends strongly on dimension \cite{Caspar:2016}.
The asymptotic finite-size scaling is given by
\eqa{\Delta \sim \begin{cases} ~ L^{-2} \ , \hspace{58pt} d = 1 \ ,\\ 
 ~ L^{-2} \, (\hsp\log L)^{-1} \ , \hspace{10pt} d = 2 \ ,\\ 
 ~ L^{-3} \ , \hspace{59pt} d = 3 \ , \end{cases}}
which is illustrated in Fig.\ \ref{Fig:DissipativeGap}b for the $d = 1$, $d = 2$ (square), and $d = 3$ (primitive cubic) lattice geometries.

\subsection{Universality of the dissipative cooling process}
\label{SubSec:Universality of the dissipative cooling process}

The observed nontrivial finite size scaling of the dissipative gap for different dimensions $d$ is remarkable. 
This concerns in particular the logarithmic correction in $d=2$.
One might speculate whether this behavior is due to the presence of topological defects (e.g., vortices) in the system.
The deeper mathematical reason for this scaling behavior and its relation to the properties of $\mathcal{M}$ has not been fully elucidated yet. 
Consequently, we checked that this scaling behavior is not an artifact of the lattice structure or the boundary conditions of the problem.

\begin{figure}[!t]
\includegraphics[width=\textwidth]{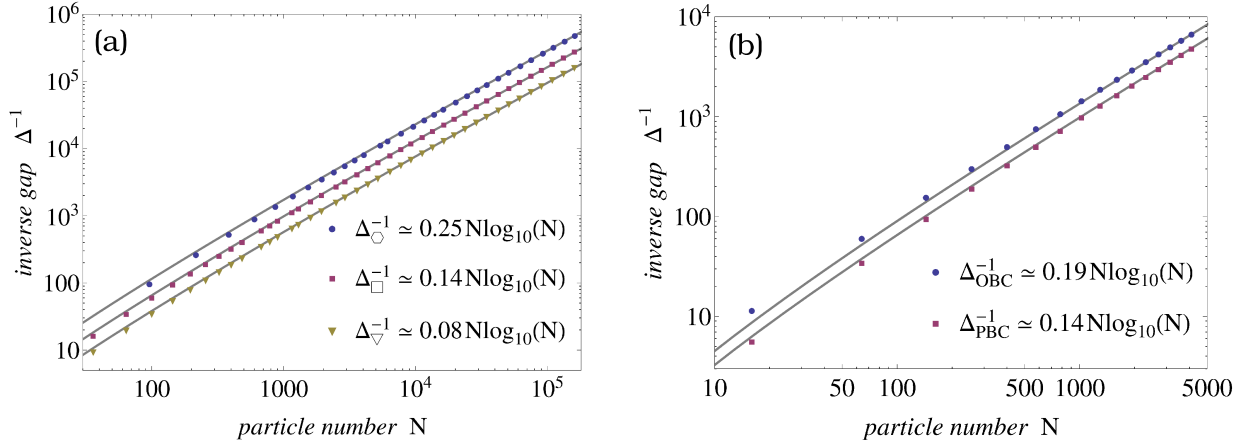}
\caption{[\textit{Color online}]
 (a) Inverse gap $\Delta^{-1}$ for different $d = 2$ dimensional lattice geometries, i.e., square (${\scriptsize\Box}$), triangular ($\triangledown$), and honeycomb ($\hexagon$), with fixed (periodic) boundary conditions as a function of the system size $N$.
 (b) Inverse gap $\Delta^{-1}$ for a $d = 2$ dimensional square lattice with open ({\it dots}) and periodic ({\it squares}) boundary conditions as a function of the system size $N$.}
\label{Fig:DissipativeGapLattice}
\end{figure}

In Fig.\ \ref{Fig:DissipativeGapLattice}a, we display the scaling of $\Delta$ for $d = 2$ dimensions upon changing the lattice geometry for fixed (periodic) boundary conditions. 
Specifically, we consider three different regular tilings of the plane, corresponding to a square (${\scriptsize\Box}$), triangular ($\triangledown$), and honeycomb ($\hexagon$) lattice geometry, respectively, and clearly observe the $\Delta^{-1} \sim N\log N$ scaling for large $N$ in all cases. 
However, note that the numerical values of $\Delta(N)$ differ.
Empirically, we find that $\Delta_{\triangledown}>\Delta_{\scriptsize\Box}>\Delta_{\scriptsize\hexagon}$ (see Fig.\ \ref{Fig:DissipativeGapLattice}a), which can be attributed to the decrease of coordination number, i.e., $n_{c,\triangledown}>n_{c,\scriptsize\Box}>n_{c,\scriptsize\hexagon}$.

Finally, we also investigated how the choice of boundary condition influences the finite-size scaling of the dissipative gap $\Delta$.
In Fig.\ \ref{Fig:DissipativeGapLattice}b, we consider a $d = 2$ system on a square lattice with either open or periodic boundary conditions.
Note that we are restricted to comparatively small lattice sizes when translation invariance is not imposed on the system.
Nevertheless, we clearly observe that the above scaling is not altered by the choice of boundary conditions and we find $\Delta_{\textrm{pbc}}>\Delta_{\textrm{obc}}$.
Summarizing, our results indicate that the dynamics of the purely dissipative cooling process for $d=2$ is rather insensitive to the lattice structure or boundary conditions, and we expect a similar behavior for $d=1$ and $d=3$.

\section{Dissipative dynamics in the presence of competing unitary and nonunitary processes}
\label{Sec:Dissipative dynamics in the presence of competing unitary and nonunitary processes}

Here, we supplement the purely dissipative cooling process considered in Sec.\ \ref{Sec:Dissipative cooling into a BEC} by competing unitary and nonunitary processes. We restrict ourselves to Hamiltonians that include only a coupling to a uniform magnetic field, which is a necessary requirement to close the hierarchy of correlation functions. Moreover, we study the effect of thermally induced spin flips via an additional nonunitary process.

\subsection{Competing unitary dynamics in the presence of a magnetic field}
\label{SubSec:Competing unitary dynamics in the presence of a magnetic field}

We consider the effect of an external field that points in the $1$-direction, as described by
\eq{H_1 = h\sum_{x}s_x^{1} \ ,}
with $h>0$ without loss of generality. 
Note that in terms of hardcore bosons, this Hamiltonian describes the equally probable creation and annihilation of particles.
Unlike the cooling operator $Q_{xy}$, this Hamiltonian does not conserve the total spin projection $S^3$.
The equations of motion for correlation functions receive additional contributions from the Hamiltonian part in Eq.\ \eqref{Eq:AdjointLindblad}, for which we need to calculate the following commutators
\seqa{\left[H_1,C_{xy}\right] &=& 2\left[H_1,F_{xy}\right] = -2\left[H_1,D_{xy}\right] = 2ihE_{xy} \ ,\\
     \left[H_1,E_{xy}\right] &=& 2ih(D_{xy}-F_{xy}) \ ,}
where the operators $E_{xy}=s_x^2s_y^3+s_x^3s_y^2$ and $F_{xy}=s_x^2s_y^2$ have been introduced, in addition to $C_{xy}$ and $D_{xy}$ to close the set of evolution equations. The dissipative contributions that arise from the action of the adjoint map $\boldsymbol{\mathcal{L}}_{xz}^{\boldsymbol{\ast}}$ are proportional to
\seqa{Q_{xz}^\dagger\big[E_{xy},Q_{xz}\big]+\big[Q_{xz}^\dagger,E_{xy}\big]Q_{xz} &=& \frac{1}{2} (1-\delta_{yz}) (E_{yz}-E_{xy})  \ ,\\
     Q_{xz}^\dagger\big[F_{xy},Q_{xz}\big]+\big[Q_{xz}^\dagger,F_{xy}\big]Q_{xz} &=& \frac{1}{2} (1-\delta_{yz}) (F_{yz}-F_{xy}) + \delta_{yz} \left( F_{xx} - D_{xy} - \frac{1}{2} C_{xy}  \right) \ . \qquad \quad  }
Proceeding along the same lines as in the previous section, we obtain the full set of evolution equations
\seqa{\label{Eq:EvolutionEquations2ptCorrelationsMagneticFieldBegin}
 \frac{d}{d\tau}\mathcal{C}_{xy} &=& \frac{1}{4}\left(\Delta_x+\Delta_y\right)\mathcal{C}_{xy}-\frac{1}{2}\delta_{\langle x,y\rangle}\left(\mathcal{C}_{xy}+4\mathcal{D}_{xy}\right)-2\eta\,\mathcal{E}_{xy} \ ,\\
 \frac{d}{d\tau}\mathcal{D}_{xy} &=& \frac{1}{4}\left(\Delta_x+\Delta_y\right)\mathcal{D}_{xy}+\frac{1}{2}\delta_{\langle x,y\rangle}\left(\mathcal{D}_{xy}-\mathcal{D}_{xx}\right)+\eta\,\mathcal{E}_{xy} ,\\
 \frac{d}{d\tau}\mathcal{E}_{xy} &=& \frac{1}{4}\left(\Delta_x+\Delta_y\right)\mathcal{E}_{xy}+\frac{1}{2}\delta_{\langle x,y\rangle}\,\mathcal{E}_{xy}+2\eta\,(\mathcal{F}_{xy}-\mathcal{D}_{xy}) \ ,\\
 \frac{d}{d\tau}\mathcal{F}_{xy} &=& \frac{1}{4}\left(\Delta_x+\Delta_y\right)\mathcal{F}_{xy}+\frac{1}{4}\delta_{\langle x,y\rangle}\left(2\mathcal{F}_{xy}-2\mathcal{D}_{xy}-\mathcal{C}_{xy}\right)-\eta\,\mathcal{E}_{xy} \ ,
 \label{Eq:EvolutionEquations2ptCorrelationsMagneticFieldEnd}}
where $\eta = h/\gamma$ is the dimensionless magnetic field variable, rescaled by the dissipative coupling $\gamma$. The diagonal elements are constant in time, $\mathcal{C}_{xx} = 4\mathcal{D}_{xx} = 4\mathcal{F}_{xx} = 1$, $\mathcal{E}_{xx} = 0$, and the system of first-order differential equations \eqref{Eq:EvolutionEquations2ptCorrelationsMagneticFieldBegin} -- \eqref{Eq:EvolutionEquations2ptCorrelationsMagneticFieldEnd} can be solved via diagonalization of the corresponding linear differential operator $\mathcal{M}_\eta$. We assume an incoherent thermal ensemble at infinite temperature as the initial state, such that $\mathcal{C}_{xy}(0) = 4\mathcal{D}_{xy}(0) = 4\mathcal{F}_{xy}(0) = \delta_{xy}$ and $\mathcal{E}_{xy}(0) = 0$.

\begin{figure}[!t]
\includegraphics[width=0.7\textwidth]{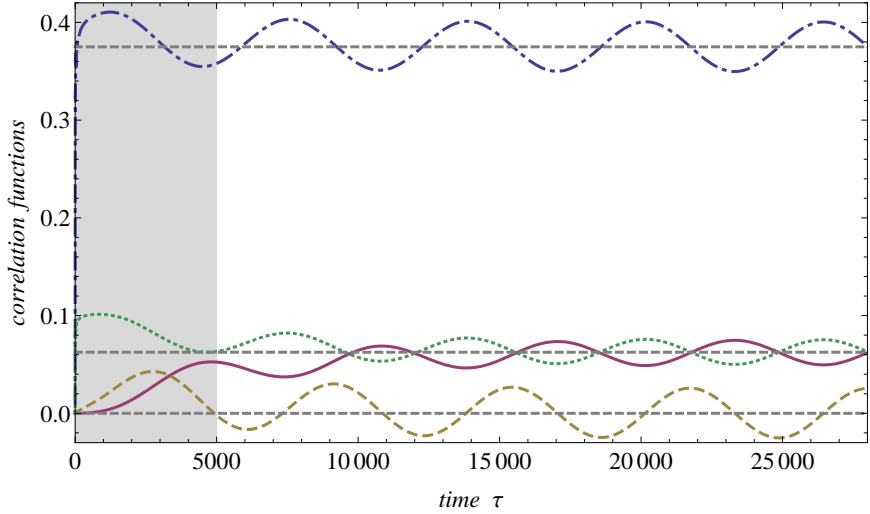}
\caption{[\textit{Color online}]
Time evolution of the two-point correlation functions $\mathcal{C}_{xy}(\tau)$ ({\it dot-dashed}), $\mathcal{D}_{xy}(\tau)$ ({\it solid}), $\mathcal{E}_{xy}(\tau)$ ({\it dashed}), $\mathcal{F}_{xy}(\tau)$ ({\it dotted}) for adjacent sites $\langle x,y\rangle$ on a $d = 2$ dimensional square lattice for $N = 4096$ and $\eta = 10^{-3}$.
The dashed horizontal lines denote late-time averages. The gray region indicates the time region for which $\tau < 1/\Delta$.}
\label{Fig:CorrelationsH1}
\end{figure}

We first investigate the asymptotic behavior of the system by studying the spectrum of the linear differential operator $\mathcal{M}_{\eta}$, which is defined by the system of Eqs.\ \eqref{Eq:EvolutionEquations2ptCorrelationsMagneticFieldBegin} -- \eqref{Eq:EvolutionEquations2ptCorrelationsMagneticFieldEnd}. For $\eta = 0$, the four largest eigenvalues of $\mathcal{M}_{\eta}$ (in decreasing order of their respective real part) are given by $\lambda_1 = \lambda_2 = \lambda_3 = 0$ and $\lambda_4 = -\Delta < 0$, where the dissipative gap $\Delta \equiv \Delta_{\eta = 0}$ is determined in Sec.\ \ref{SubSec:Dynamics of purely dissipative cooling}. Switching on the external magnetic field $\eta$, two of the zero eigenvalues pick up an imaginary part, by which the spectrum is modified as follows: $\lambda_1 = 0$, $\lambda_{2,3} = \pm 2 i\eta$, and $\lambda_4 = -\Delta$. This behavior has immediate consequences.
First, the longest timescale in the system, which is determined by the dissipative gap, is not changed in the presence of an external magnetic field.
On the other hand, the imaginary eigenvalues $\pm 2 i\eta$ indicate that the system does not converge to a unique final steady state. 
The correlations are rather seen to exhibit oscillations with frequency $2\eta$ around an average asymptotic value -- the system is asymptotically characterized by a limit cycle.
We determine the late-time behavior analytically for $\tau \gg 1/\Delta$ from Eqs.\ \eqref{Eq:EvolutionEquations2ptCorrelationsMagneticFieldBegin} -- \eqref{Eq:EvolutionEquations2ptCorrelationsMagneticFieldEnd}:
\seqa{\label{Eq:LateTimeBehaviorMagneticField}
 \mathcal{C}_{xy}(\tau\gg 1/\Delta) &=& \frac{3+5\delta_{xy}}{8}+2(1-\delta_{xy})g(\eta,N)\cos(2\eta\tau- \varphi) \ ,\\
 \mathcal{D}_{xy}(\tau\gg 1/\Delta) &=& \frac{1+3\delta_{xy}}{16}-(1-\delta_{xy})g(\eta,N)\cos(2\eta\tau - \varphi) \ ,\\
 \mathcal{E}_{xy}(\tau\gg 1/\Delta) &=& 2(1-\delta_{xy})g(\eta,N)\sin(2\eta\tau - \varphi) \ ,\\
 \mathcal{F}_{xy}(\tau\gg 1/\Delta) &=& \frac{1+3\delta_{xy}}{16}+(1-\delta_{xy})g(\eta,N)\cos(2\eta\tau - \varphi) \ .}
Here, $\varphi$ is an irrelevant phase offset and the function $g(\eta,N)$ describes the oscillation amplitude at late times. 
In Fig.\ \ref{Fig:CorrelationsH1} we display the time evolution of the various two-point correlation functions, which clearly exhibits the oscillatory behavior at late times.
Moreover, we numerically determine $g(\eta,N)$ for which we observe a monotonic decay with increasing magnetic field strength $\eta$, where $g(\eta\to 0,N) = 1/16$ and $g(\eta\to \infty,N) = (8N\eta)^{-1}$.

\begin{figure}[!t]
\includegraphics[width=0.7\textwidth]{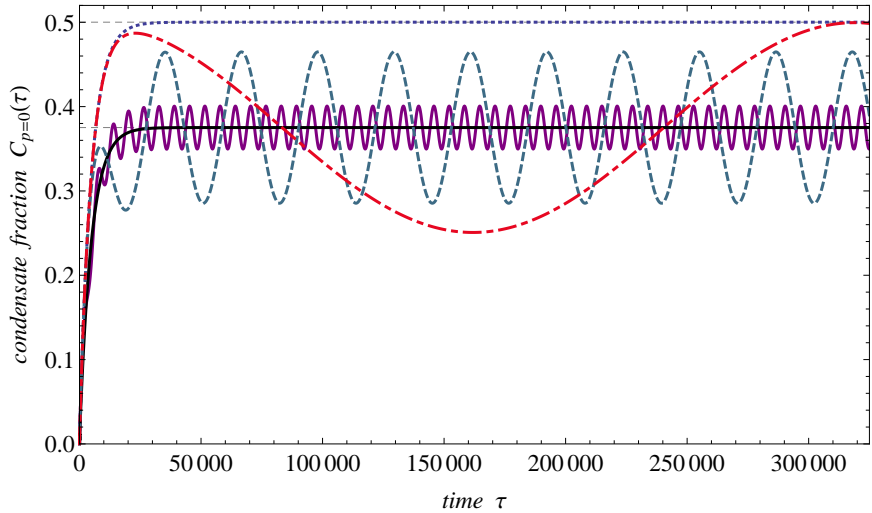}
\caption{[\textit{Color online}]
Time evolution of the condensate fraction $\mathcal{C}_{p=0}(\tau)$ for $N = 4096$ on a $d = 2$ dimensional square lattice for different external magnetic fields $\eta = 0$ ({\it dotted}), $\eta = 1$ ({\it solid, black}), $\eta = 5\cdot 10^{-4}$ ({\it solid, purple}), $\eta = 10^{-4}$ ({\it dashed}) and $\eta = 10^{-5}$ ({\it dot-dashed}).
The dashed horizontal lines indicate the asymptotic value $\mathcal{C}_{p=0}^{\infty} = (N+1)/(2N)$ as well as the late-time average $\lim_{\tau\rightarrow \infty} \frac{1}{\tau} \int_{\Delta^{-1}}^{\tau} \textrm{d}\tau' \, {\mathcal{C}}_{p=0}(\tau') = (3N+5)/(8N)$.}
\label{Fig:CondensateH1}
\end{figure}

Finally, we display the time evolution of the condensate fraction $\mathcal{C}_{p=0}(\tau)$ for different values of $\eta$ in Fig.\ \ref{Fig:CondensateH1}.
According to \eqref{Eq:LateTimeBehaviorMagneticField}, its late-time behavior is determined by
\eq{\label{Eq:ZeroModeMagneticField}\mathcal{C}_{p=0}(\tau\gg 1/\Delta) = \frac{3N+5}{8N}+\frac{2(N-1)}{N}g(\eta,N)\cos(2\eta\tau-\varphi_0) \ .}
In comparison to the purely dissipative cooling dynamics ($\eta = 0$, cf.\ Sec.\ \ref{Sec:Dissipative cooling into a BEC}) for which the steady-state condensate fraction $\mathcal{C}_{p=0}^{\infty} = (N+1)/(2N)$ is approached, we find that the presence of a nonvanishing magnetic field substantially decreases the late-time average value 
\eq{\lim_{\tau\rightarrow \infty} \frac{1}{\tau} \int_{\Delta^{-1}}^{\tau} \textrm{d}\tau' \, {\mathcal{C}}_{p=0}(\tau') = \frac{3N+5}{8N} \ .}

\subsubsection{Model two-spin system}
\label{SubSubSec:Model two-spin system}

To understand the main features of the dissipative real-time dynamics of Eqs.\ \eqref{Eq:LateTimeBehaviorMagneticField} and \eqref{Eq:ZeroModeMagneticField}, we may consider a model system consisting of two particles for which the time evolution of the density matrix can be easily calculated explicitly.
To this end, we provide the initial density matrix
\eq{\rho(0) = \frac{1}{4}\left(\ket{t_+}\bra{t_+}+\ket{t_-}\bra{t_-}+\ket{t_0}\bra{t_0}+\ket{s}\bra{s}\right) \ ,}
with the triplet states $\ket{t_+} = \ket{\uparrow\uparrow}$, $\ket{t_-} = \ket{\downarrow\downarrow}$, $\ket{t_0} = (\ket{\uparrow\downarrow}+\ket{\downarrow\uparrow})/\sqrt{2}$, as well as the singlet state $\ket{s} = (\ket{\uparrow\downarrow}-\ket{\downarrow\uparrow})/\sqrt{2}$.
For vanishing magnetic field, the time-dependent density matrix is given by
\eq{\rho^{\eta=0}(\tau) = \frac{1}{4}\left(\ket{t_+}\bra{t_+}+\ket{t_-}\bra{t_-}+(2-e^{-\tau})\ket{t_0}\bra{t_0}+e^{-\tau}\ket{s}\bra{s}\right) \ ,}
i.e., the singlet state $\ket{s}\bra{s}$ is mapped to the triplet state $\ket{t_0}\bra{t_0}$ while the other components remain invariant under the time evolution. This picture changes in the presence of a nonvanishing magnetic field. While the dissipative process of Sec.\ \ref{Sec:Dissipative cooling into a BEC} still eliminates the singlet component from the ensemble, the magnetic field results in a mixing of the three triplet states
\eqa{\rho^{\eta}(\tau) &=& \frac{1}{4}e^{-\tau}\ket{s}\bra{s}+\frac{c_1(\tau)}{16(1+4\eta^2)}\left(\ket{t_+}\bra{t_+}+\ket{t_-}\bra{t_-}\right) \nonumber \\
 && +\: \frac{c_2(\tau)}{8(1+4\eta^2)}\ket{t_0}\bra{t_0}+\frac{c_3(\tau)}{16(1+4\eta^2)}\left(\ket{t_+}\bra{t_-}+\ket{t_-}\bra{t_+}\right) \nonumber \\
 && +\: \frac{i c_4(\tau)}{8\sqrt{2}(1+4\eta^2)}\left(\ket{t_+}\bra{t_0}+\ket{t_-}\bra{t_0}-\ket{t_0}\bra{t_+}-\ket{t_0}\bra{t_-}\right) \ ,}
where the time-dependent coefficients are given by
\seqa{c_1(\tau) &=& 5(1+4\eta^2)-4\eta^2e^{-\tau}-\cos(2\eta \tau)-2\eta \sin(2\eta \tau) \ ,\\
 c_2(\tau) &=& 3(1+4\eta^2)-2(1+2\eta^2)e^{-\tau}+\cos(2\eta \tau)+2\eta \sin(2\eta \tau) \ ,\\
 c_3(\tau) &=& 1+4\eta^2-4\eta^2e^{-\tau}-\cos(2\eta \tau)-2\eta \sin(2\eta \tau) \ ,\\
 c_4(\tau) &=& -2\eta e^{-\tau}+2\eta\cos(2\eta \tau)-\sin(2\eta \tau) \ .}
Defining the two-particle operator $C = s^+\otimes \hsp s^-+s^-\otimes \hsp s^+$, we evaluate $C\ket{t_+} = C\ket{t_-} = 0$, $C\ket{t_0} = \ket{t_0}$, and $C\ket{s} = -\ket{s}$.
Thus, it is sufficient to consider only the diagonal elements of the density matrix, denoted by $\rho_{\textrm{diag}}$, to calculate the expectation value $\mathcal{C} = \operatorname{tr} \{ \rho \hsp C \}$. In fact, at late times, the time-averaged diagonal entries $\rho_{\textrm{diag}}$ are given by
\eq{\lim_{\tau\to\infty} \frac{1}{\tau} \int^{\tau}_{\Delta^{-1}} \textrm{d}\tau' \, \rho^{\eta}_{\textrm{diag}}(\tau') = \frac{1}{16}\left(5\ket{t_+}\bra{t_+}+5\ket{t_-}\bra{t_-}+6\ket{t_0}\bra{t_0}\right) \ .}
This result should be compared to the nonequilibrium steady-state density matrix in the absence of an external field
\eq{\rho_{\textrm{NESS}}^{\eta = 0} = \frac{1}{16}\left(4\ket{t_+}\bra{t_+}+4\ket{t_-}\bra{t_-}+8\ket{t_0}\bra{t_0}\right) \ .}
Thus, the magnetic field generates an average spin rotation, which decreases the $\ket{t_0}\bra{t_0}$ contribution when compared to the final state of the driven, purely dissipative system.
As a consequence, the late-time average value $\lim_{\tau\to\infty} \frac{1}{\tau} \int^{\tau}_{\Delta^{-1}} \textrm{d}\tau' \, \mathcal{C}^{\eta}(\tau') = 3/8$ is smaller than the asymptotic value for the purely dissipative cooling dynamics $\mathcal{C}^{\eta=0} = 1/2$. 
It therefore appears that the phenomenology of the macroscopic $N$-particle system in the presence of a magnetic field in the $1$-direction is essentially captured by the corresponding two-spin model system.

\subsection{Competing thermal noise}
\label{SubSec:Competing thermal noise}

Here, we introduce the effect of an external magnetic field that points in the $3$-direction
\eq{H_3 = h\sum_{x}s_x^{3} \ ,}
where we assume that $h>0$.
In contrast to $H_1$, the Hamiltonian $H_3$ commutes with $Q_{xy}$ and therefore it does not lead to an additional coupling to two-point operators (as we encountered in the previous section).
Here, however, we allow for local spin flips via additional jump processes that are accounted for by local operators $L_{x}^{\pm} = s_{x}^{\pm}$ \cite{Prosen:2008,Wichterich:2007,Prosen:2008b,Crawford:2015}.
In general, we may assign independent interaction rates $\gamma^\pm_x$ to both processes $L_{x}^{\pm}$.
Assuming a thermal occupation of the bath, however, the spin flip rates are related via the Boltzmann factor
\eq{\frac{\gamma^+}{\gamma^-} = \exp\left(-\frac{2h}{T}\right) \ ,}
where $T$ denotes the bath temperature and we assume spatial homogeneity ($\gamma^{\pm}_x = \gamma^{\pm}$).
This relation does not set the overall interaction rate which we denote by $\kappa$. We assign the following values to the ratios 
\eq{\frac{\gamma^+}{\kappa} = n_T\ , \qquad \frac{\gamma^-}{\kappa} = n_T + 1 \ ,}
where $n_T \equiv \left(e^{2h/T}-1\right)^{-1}$ is the thermal occupation number.
The equations of motion for correlation functions \eqref{Eq:AdjointLindblad} receive additional contributions from the spin flip processes $L_x^{\pm}$:
\seqa{\big(L_x^{\pm}\big)^{\dagger}\big[s_{x}^{3},L_x^\pm\big]+\big[\big(L_x^{\pm}\big)^{\dagger},s_{x}^{3}\big]L_x^{\pm} &=& -2s_{x}^3\pm\mathbbm{1} \ ,\\
     \big(L_x^{\pm}\big)^{\dagger}\big[C_{xy},L_x^\pm\big]+\big[\big(L_x^{\pm}\big)^{\dagger},C_{xy}\big]L_x^{\pm} &=& -C_{xy} \ ,\\
     \big(L_x^{\pm}\big)^{\dagger}\big[D_{xy},L_x^\pm\big]+\big[\big(L_x^{\pm}\big)^{\dagger},D_{xy}\big]L_x^{\pm} &=& -2D_{xy}\pm s_{y}^3 \ , } 
while the commutators $[H_3,C_{xy}] = [H_3,D_{xy}] = [H_3,s_x^3]=0$ yield no additional terms.
Accordingly, the closed set of time evolution equations reads
\seqa{\label{Eq:EvolutionEquationThermalBegin}
 \frac{d}{d\tau}{\scriptstyle\mathcal{S}}_x^3 &=& \frac{1}{4}\Delta_x{\scriptstyle\mathcal{S}}_x^3-\frac{\kappa}{\gamma}(2n_T+1){\scriptstyle\mathcal{S}}_x^3-\frac{\kappa}{2\gamma} \ ,\\
 \frac{d}{d\tau}\mathcal{C}_{xy} &=& \frac{1}{4}\left(\Delta_x+\Delta_y\right)\mathcal{C}_{xy}-\frac{1}{2}\delta_{\langle x,y\rangle}\left(\mathcal{C}_{xy}+4\mathcal{D}_{xy}\right)-\frac{\kappa}{\gamma}(2n_T+1)\mathcal{C}_{xy} \ ,\\
 \frac{d}{d\tau}\mathcal{D}_{xy} &=& \frac{1}{4}\left(\Delta_x+\Delta_y\right)\mathcal{D}_{xy}+\frac{1}{2}\delta_{\langle x,y\rangle}\left(\mathcal{D}_{xy}-\mathcal{D}_{xx}\right)-\frac{2\kappa}{\gamma}(2n_T+1)\mathcal{D}_{xy}-\frac{\kappa}{2\gamma}\left({\scriptstyle\mathcal{S}}_x^3+{\scriptstyle\mathcal{S}}_y^3\right) \ , \qquad
 \label{Eq:EvolutionEquationThermalEnd}}
while $\mathcal{C}_{xx} = 4\mathcal{D}_{xx} = 1$.

The behavior of this system, which is driven by the operators $\sqrt{\gamma^{\pm}} L_x^{\pm}$ and $\sqrt{\gamma} Q_{xy}$ can be fully characterized in terms of two independent parameters -- the ratio of couplings $\kappa/\gamma$ and the effective temperature $T/h$. 
To understand the relevant modes that determine the late-time behavior, we inspect the spectrum of the linear differential operator $\mathcal{M}_T$ as defined by the system of linear equations \eqref{Eq:EvolutionEquationThermalBegin} -- \eqref{Eq:EvolutionEquationThermalEnd}. Again, we consider an incoherent infinite-temperature ensemble as initial state, for which $\mathcal{C}_{xy}(0) = 4\mathcal{D}_{xy}(0) = \delta_{xy}$ and ${\scriptstyle\mathcal{S}}_{x}^{3}(0) = 0$.

\begin{figure}[!t]
\includegraphics[width=0.7\textwidth]{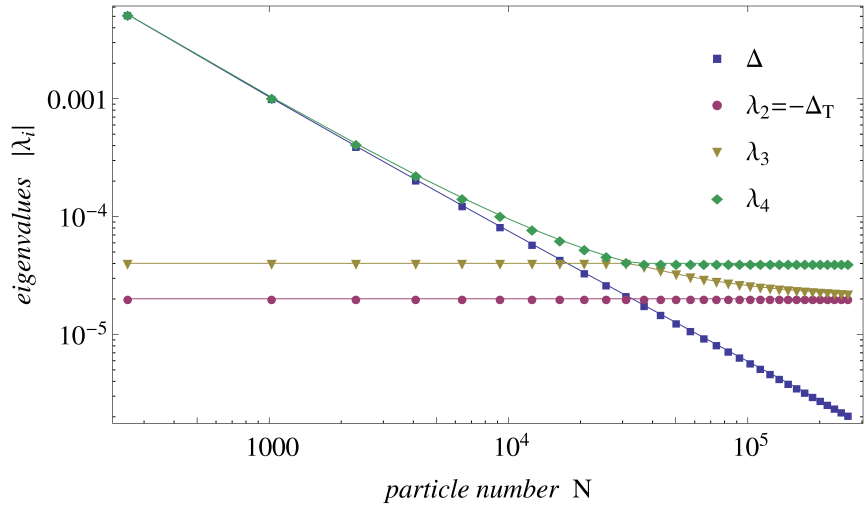}
\caption{[\textit{Color online}]
 Comparison of the dissipative gap for purely dissipative cooling dynamics $\Delta$ with the three largest nonvanishing eigenvalues $\lambda_2=-\Delta_T$, $\lambda_3$ and $\lambda_4$ of the linear differential operator $\mathcal{M}_T$ on a double-logarithmic scale as a function of the system size $N$.
 The data was obtained on a $d = 2$ dimensional lattice with parameters $\kappa/\gamma = 10^{-7}$ and $n_T = 10^2$.}
\label{Fig:SpectrumH3}
\end{figure}

The spectrum of $\mathcal{M}_T$ is real and nonpositive so that the asymptotic behavior is governed by the dissipative gap $\Delta_T$. It is independent of the system size
\eq{\Delta_T = \frac{\kappa}{\gamma}(2n_T+1) \ ,}
which is in stark contrast to purely dissipative cooling dynamics ($\kappa = 0$), for which the dissipative gap $\Delta$ exhibits a nontrivial finite-size scaling.
This means that the presence of a thermal bath dominates the asymptotic behavior of the system. More specifically, for the purely dissipative process governed by the jump operators $Q_{xy}$, the three largest eigenvalues are given by $\lambda_1 = \lambda_2 = 0$ and $\lambda_3 = -\Delta$.
The presence of thermal noise, however, lifts the degeneracy of the zero eigenvalues such that $\lambda_1 = 0$, $\lambda_2 = -\Delta_T$, and $\lambda_3 < \lambda_2$. 
In Fig.\ \ref{Fig:SpectrumH3} we compare the spectrum of a dissipative system in both cases -- with and without thermal noise -- and we display the scaling of the largest eigenvalues as a function of $N$.
We clearly observe that $\Delta_T$ is independent of the system size when compared to $\Delta$ (with $\kappa = 0$), which exhibits the nontrivial $\Delta^{-1} \sim N\log N$ scaling in $d=2$ dimensions.

\begin{figure}[!t]
\includegraphics[width=0.7\textwidth]{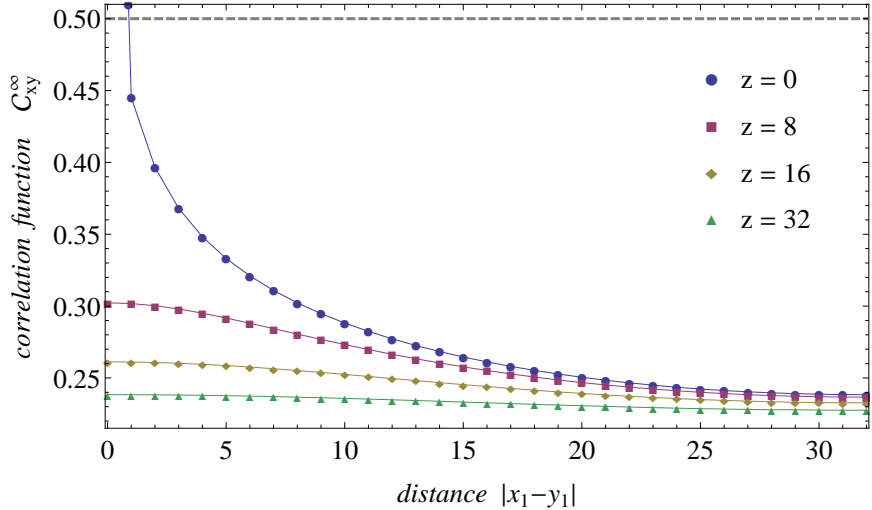}
\caption{[\textit{Color online}]
 Spatial dependence of the correlation function $\mathcal{C}_{xy}^{\infty}$ for $N = 4096$ particles on a $d = 2$ dimensional lattice for $\kappa/\gamma = 10^{-5}$ and $n_T = 10$.
 The abscissa denotes the distance in the $1$-direction $|x_1-y_1|$ while the different curves correspond to different values of the separation $z = |x_2-y_2|$.
 The dashed horizontal line denotes the asymptotic value for dissipative cooling without thermal noise, $\mathcal{C}_{xy}^{\infty} = 1/2$, for $x\neq y$.}
\label{Fig:CorrelationH3}
\end{figure}

Owing to the fact that $L_x^{\pm}$ does not conserve spin, $[s_x^3,L_x^{\pm}] \neq 0$, we obtain a nonvanishing value both for the asymptotic magnetization ${\scriptstyle\mathcal{S}}_x^3$ and for the two-point function $\mathcal{D}_{xy}$: 
\eqa{\label{Eq:AsymptoticValuesThermal}
  \lim_{\tau\to\infty}{\scriptstyle\mathcal{S}}_x^3(\tau) &=& -\frac{1}{2(2n_T+1)} \ ,\\
  \lim_{\tau\to\infty}\mathcal{D}_{xy}(\tau) &=& \frac{1}{4}\left(\delta_{xy}+\frac{1-\delta_{xy}}{(2n_T+1)^2}\right) \ .}
While these values can be calculated easily by hand, to determine $\mathcal{C}_{xy}^{\infty}$ we need to invert the linear differential operator corresponding to the linear system Eqs.\ \eqref{Eq:EvolutionEquationThermalBegin} -- \eqref{Eq:EvolutionEquationThermalEnd}, which for large system sizes, can only be done numerically. 
We observe that the correlation function $\mathcal{C}^{\infty}_{xy}$ ($\kappa \neq 0$) decays with the separation $|x-y|$ in the presence of a thermal coupling, as shown in Fig.\ \ref{Fig:CorrelationH3}. 
This is in contrast to the dissipative process governed by the operator $Q_{xy}$, for which the asymptotic value is given by $\mathcal{C}_{xy}^{\infty} = 1/2$, for $x\neq y$.
We conclude that the thermally induced spin flips counteract the cooling process and destroy the long-range order in the system, thereby introducing a correlation length $\xi\ll L$.

\begin{figure}[t!]
\includegraphics[width=\textwidth]{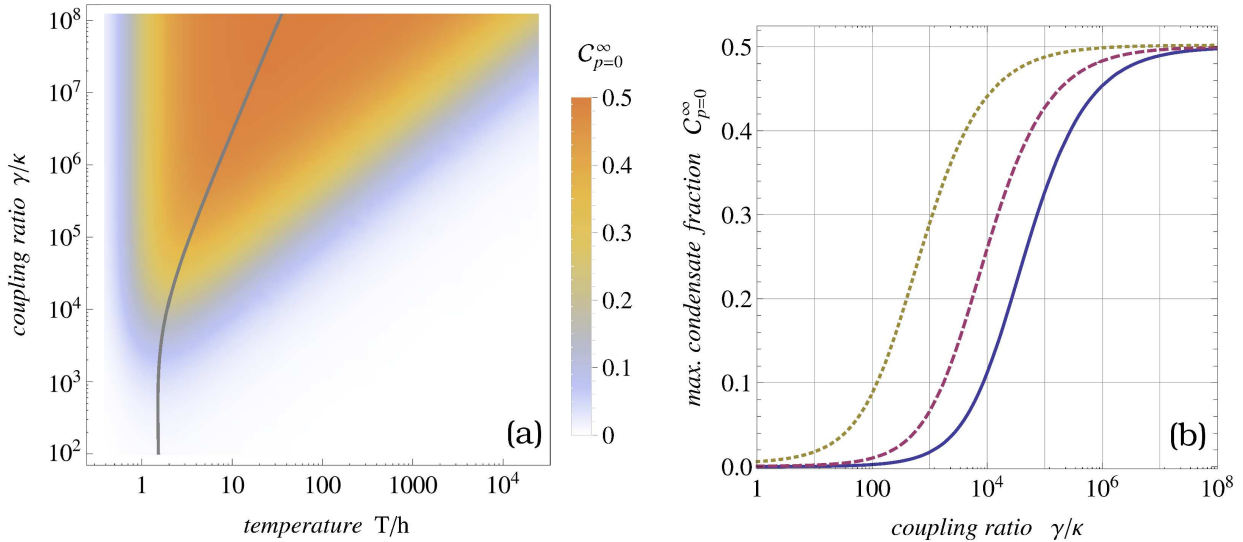}
\caption{[\textit{Color online}]
 (a) Asymptotic condensate fraction $\mathcal{C}_{p=0}^{\infty}$ for $N = 4096$ particles on a $d=2$ dimensional lattice as a function of $\gamma/\kappa$ and $T/h$. 
 The gray line indicates the location of the maximum value of $\mathcal{C}_{p=0}^{\infty}$ for given $\gamma/\kappa$.
 (b) Maximum value of $\mathcal{C}_{p=0}^{\infty}$ as a function of the ratio $\gamma/\kappa$ for $N=9216$ ({\it solid}), $N=2304$ ({\it dashed}) and $N=256$ ({\it dotted}).}
\label{Fig:CondensateH3}
\end{figure}

The asymptotic value of the zero mode $\mathcal{C}_{p=0}^{\infty}$ strongly depends on the parameters $\gamma/\kappa$ and $T/h$, as well as the particle number $N$.
Setting the number of particles to $N=4096$, we calculate the nonequilibrium steady-state phase diagram, which is shown in Fig.\ \ref{Fig:CondensateH3}a. In this case, we find that the thermal noise destroys the long-range order more or less completely in the range $\gamma/\kappa\lesssim\mathcal{O}(10^3)$, independent of the temperature $T$.
This means that the coupling of the cooling process $\gamma$ needs to be larger than that of the thermal bath $\kappa$ by several orders of magnitude in order to generate a macroscopically ordered state. 
On the other hand, for large values of $\gamma/\kappa$, we observe an intriguing dependence of $\mathcal{C}_{p=0}^{\infty}$ on the temperature $T$ [cf.\ Fig.\ \ref{Fig:CondensateH3}a]. 
That is, for fixed $\gamma/\kappa$ and finite $N$, we find a non-monotonous behavior of $\mathcal{C}_{p=0}^{\infty}(T)$, where in the limiting cases $\lim_{T\to 0}\mathcal{C}_{p=0}^{\infty}(T) = \lim_{T\to\infty}\mathcal{C}_{p=0}^{\infty}(T) = 1/N$. In fact, these values corresponds to the lower bound for the zero mode $\mathcal{C}_{p = 0}$.
This can be understood in the following way: In the limit $T\to0$, spin-flip operations $s_x^+$ are forbidden, such that the action of $s_x^-$ on all possible sites aligns all spins along the negative $3$-direction, cf.\ Eq.\ \eqref{Eq:AsymptoticValuesThermal}, hence destroying any long-range order in the $(1,2)$-plane. 
In the opposite limit $T\to\infty$, we have $\gamma^+=\gamma^-\to\infty$ such that the spin-flip operators $s_x^+$ and $s_x^-$ completely dominate the dynamics and, hence, prohibit any long-range order.
Between these two limits, for any value of $\gamma/\kappa$, we observe that there exists an optimal value $T_{\textrm{opt}}$ for which $\mathcal{C}_{p=0}^\infty$ takes its maximum value [cf.\ Fig.\ \ref{Fig:CondensateH3}b], also indicated by the continuous line in Fig.\ \ref{Fig:CondensateH3}a.
Thus, we may find a distinct temperature region $T_1 < T < T_2$ in which the engineered dissipation performs robustly even in the presence of thermal noise. 

This region depends on the ratio of couplings $\gamma/\kappa$ and the system size $N$.
As the number of particles $N$ is increased, the nearest-neighbor symmetrizing action $Q_{xy}$ finds it more and more difficult to compete against the thermal coupling that acts locally.
Thus, we expect that in the thermodynamic limit $N\to\infty$, single spin flips eventually destroy any long-range order and therefore dominate for any finite value of $\gamma/\kappa$.
In contrast, for a small number of particles $N$ the value of $\gamma/\kappa$ that is required to generate long-range order decreases as well [cf.\ Fig.\ \ref{Fig:CondensateH3}b]. 
In view of experimental realizations \cite{Schindler:2013}, our findings indicate that thermal fluctuations are not too prohibitive for generating long-range order via engineered dissipation, at least for not too large systems.

\section{Conclusions}
\label{Sec:Conclusions}

In this work we have studied the time evolution of correlations in the context of an open Markovian quantum many-body system, which was originally proposed for the dissipative cooling into a BEC \cite{Diehl:2008,Kraus:2008,Schindler:2013}. 
In a previous publication \cite{Caspar:2016} we showed that the corresponding purely dissipative process governed by the single non-Hermitian quantum jump operator $Q_{xy} = \frac{1}{2} \left( s_x^{+} + s_y^{+} \right) \left( s_x^{-} - s_y^{-} \right)$ allows for a semi-analytic solution of spin-spin correlation functions. Here, we have extended these results by studying the universality of the dissipative process.

We have established that the novel finite-size scaling behavior of the dissipative gap is in fact insensitive to the choice of lattice discretization as, e.g., provided by the lattice geometry or boundary conditions.
Furthermore, we have studied the stability of the dissipative cooling process with respect to unitary and nonunitary perturbations.
To this end, we derived conditions under which additional perturbations can be considered within the framework of a closed hierarchy of correlation functions, thereby admitting a closed analytic solution to the nonequilibrium dynamics.
In particular, we allowed for the presence of a uniform magnetic field, as well as a coupling to a thermal bath $\kappa$, while the system is driven by the dissipative cooling process with a uniform rate $\gamma$. 
We calculated the nonequilibrium steady-state phase diagram and found that for any finite particle number $N$, above a certain threshold value for the coupling ratio $\gamma/\kappa$, the system allows for a final state $\rho_{\textrm{NESS}}$ with long-range order.
However, the efficiency of the dissipative cooling mechanism decreases with the system size $N$ and the correlation length eventually becomes zero in the thermodynamic limit $N\rightarrow \infty$.
We provided concrete numerical bounds for the dissipative couplings $\gamma$ and $\kappa$, as well as the system size $N$ in order for the dissipative process to perform robustly.

The following picture appears with regard to experiments: It seems that the finite system size $N$ is crucial for the dissipative cooling protocol to generate macroscopic order in the presence of a nonvanishing coupling to a thermal bath. 
That is, our results indicate that the driving with jump operators $Q_{xy}$ can be competitive only for small $N$, where the necessary ratio of dissipative couplings $\gamma/\kappa$ to achieve long-range correlations is small. 
These results are consistent with a recent experimental realization of the dissipative cooling protocol using up to $N = 4$ particles \cite{Schindler:2013} and it is reasonable to expect that similar proposals for dissipative state generation might face the same limiting constraints with respect to system size.

It is a natural question whether our discussion of closed hierarchies for $s=1/2$ quantum spin systems can be generalized to arbitrary spin representations. This would provide a unique means of studying the classical limit $(s\rightarrow \infty)$ of driven open quantum spin systems and possibly other types of dissipative dynamics with distinct properties of the final state. Additional information on the asymptotic dynamics for quantum dissipative processes can be obtained by investigating the linear response \cite{Kubo:1957mj} in the vicinity of the nonequilibrium steady state. 
This would allow us to inquire about the presence of generalized fluctuation-dissipation relations (see, e.g., Refs.\ \cite{Kubo:1966,Albert:2015,CamposVenuti:2016}) in the general setting of open Markovian quantum dynamics. We hope to address these questions in future work.

\section{Acknowledgments}
\label{Sec:Acknowledgments}
We thank J.\ I.\ Cirac, S.\ Diehl, C.\ Tretter, and P.\ Zoller for illuminating discussions.
This research is funded by the European Research Council under the European Union's Seventh Framework Programme (FP7/2007-2013)/ERC under grant agreement 339220.

\end{document}